\documentclass[twocolumn,secnumarabic,amssymb, nobibnotes, aps,prd]{revtex4} 

\usepackage{graphicx,amssymb,amsmath,dcolumn,bm}
\begin{document}

\title{Bistability double crossing curve effect in a polariton-laser semiconductor microcavity}

\author{ E. A. Cotta}
\email{cotta@fisica.ufmg.br}
\author{ F. M. Matinaga}

\affiliation{Departamento de Fisica, Universidade Federal de Minas
Gerais, Belo Horizonte, Minas Gerais, Brazil}

\date{\today}

\begin{abstract}
We report an experimental observation of polaritonic optical
bistability of the laser emission in a planar semiconductor
microcavity with a $100\AA$ GaAs single quantum well in the
strong-coupling regime. The bistability curves show crossings that
indicate a competition between a Kerr-like effect induced by the
polariton population and thermal effects. Associated with the
bistability, laser-like emission occurs at the bare cavity mode.
\end{abstract}


\maketitle

Bistability is a very general phenomenon involving a phase
transition far from thermal equilibrium. Since its first observation
in a passive, unexcited sodium vapor medium\cite{Gibbs}, it has been
observed in many different materials including tiny semiconductor
etalons. In counterparts, many of the phenomena studied in lasers,
such as fluctuations, regenerative pulsations, and optical
turbulence, can be observed in passive bistable systems, often under
better controlled conditions. Their applications are based in
dispersive and temperature switching using optical bistability (OB)
properties, suggesting possibilities in high-speed all-optical
signal processing.

Generally an optical bistability (OB) event requires that both the
input and the output are optical signals. In order to obtain OB,
light and matter must be closely coupled together. The cavity field,
which is enhanced by the resonance condition, depends strongly on
the dielectric function of the medium inside the cavity producing
the optical feedback. The sharpness of the exciton and cavity modes
results in the strong coupling. This means that the exciton mode and
the cavity photon mode are in such close interaction as to give rise
to two polariton states resulting in an anti-crossing
(Rabi-splitting) between the upper and lower polariton branches.

The scattering mechanism is marked by generation of two polaritons
with wave-vectors $\overrightarrow{k} = 0$ (signal) and
$\overrightarrow{k} = 2\overrightarrow{k_p}$ (idler), where
$\overrightarrow{k_p}$ is the excitation wave-vector, which can be
controlled by the incidence angle and corresponds to the nontrivial
solution for the energy conservation condition. In this case, we
have a better optimization of this parametric amplification and a
minimum in the parametric scattering threshold intensity. Thus, the
``magic angle'' is $\theta_m =
arcsen(n_{qw}\sqrt{\Omega/\omega_{e-p}})\approx 9^o \pm 2^o$, for a
$\lambda$-cavity containing a single quantum well like gain media
with refractive index $n_{qw}$, Rabi-splitting energy of
$\hbar\Omega = 3.4meV$ and a exciton-polariton (e-p) peak emission
in $\omega_{e-p}$.

The nonlinearities associated with the reduced normal mode
splitting, while the excitation power is increased, produce a
collapse of the polariton modes and the crossover between the strong
and weak regimes\cite{Houdre, Lygnes, Butte}.  These effects have
been interpreted in terms of bleaching of the exciton
absorption\cite{Houdre} and excitation induced
dephasing\cite{Lygnes} due to exciton-exciton interaction.
Therefore, all the nonlinear behavior would affect strongly the
optical response of the microcavity in a resonant excitation
condition. Thus, we report an experimental study of the optical
nonlinearities of a normal mode in a semiconductor microcavity by
means of laser emission, demonstrating the presence of a bistable
response under resonant excitation conditions. We investigate the
optical bistability providing a detuning between cavity-exciton and
pump laser-exciton energies which play an important role in the
nonlinear kinetics of microcavity polariton scattering as well.

Through this resonant excitation conditions, Ciuti \emph{et al.}
define an effective Hamiltonian\cite{Ciuti} for polariton-polariton
interaction analogous to the Hamiltonian for an optical Kerr medium.
The difference here is the dependence of the refractive index on the
polariton number instead of the photon number. This results in
bistable behavior for high enough excitation intensities, as in the
Kerr medium inside a cavity. OB has been observed in this manner in
both the weak\cite{Sfez} and strong-coupling\cite{Cavigli} regimes
(under different conditions).

Moreover, due to a large variety of nonlinearities in microcavity
normal modes, the pump power heats the sample proportionally to the
increase of the absorption at resonance. This may results in a
bistable behavior for high enough excitation intensities, as in a
Kerr medium in a cavity, producing different kinds of OBs.

In a recent work\cite{Giacobino}, the dynamic equations for the
polariton field and the bistability threshold were analyzed, as well
as the reflectivity and transmission spectra, but still in
steady-state regime. Following these theoretical results, we realize
an experimental observation of the OB caused by changes in the
refractive index due to a large e-p population in a semiconductor
microcavity in the strong-coupling regime through e-p laser
emission. The behavior of the bistable response, varying the ``cw''
pump energy as well as the cavity resonance, reveals singularities
such as double crossings between the branches of the OB curve, which
play an important role in the nonlinear kinetics of microcavity
polariton scattering. An interpretation of the data is presented,
but a detailed theoretical explanation of these observations is
beyond the scope of this work.

\begin{figure}[htbp!]
  \includegraphics[height=4cm]{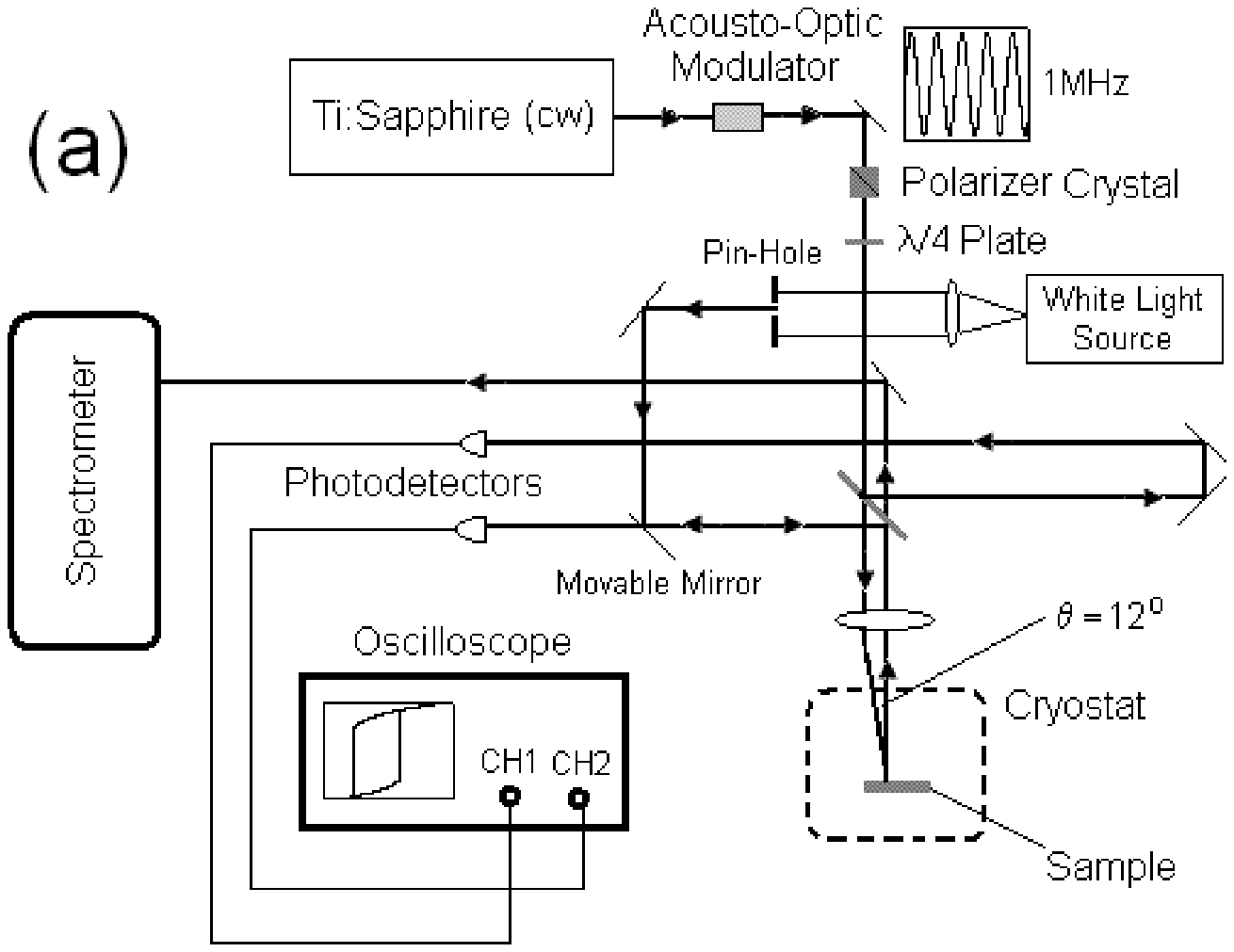}
  \includegraphics[height=4cm]{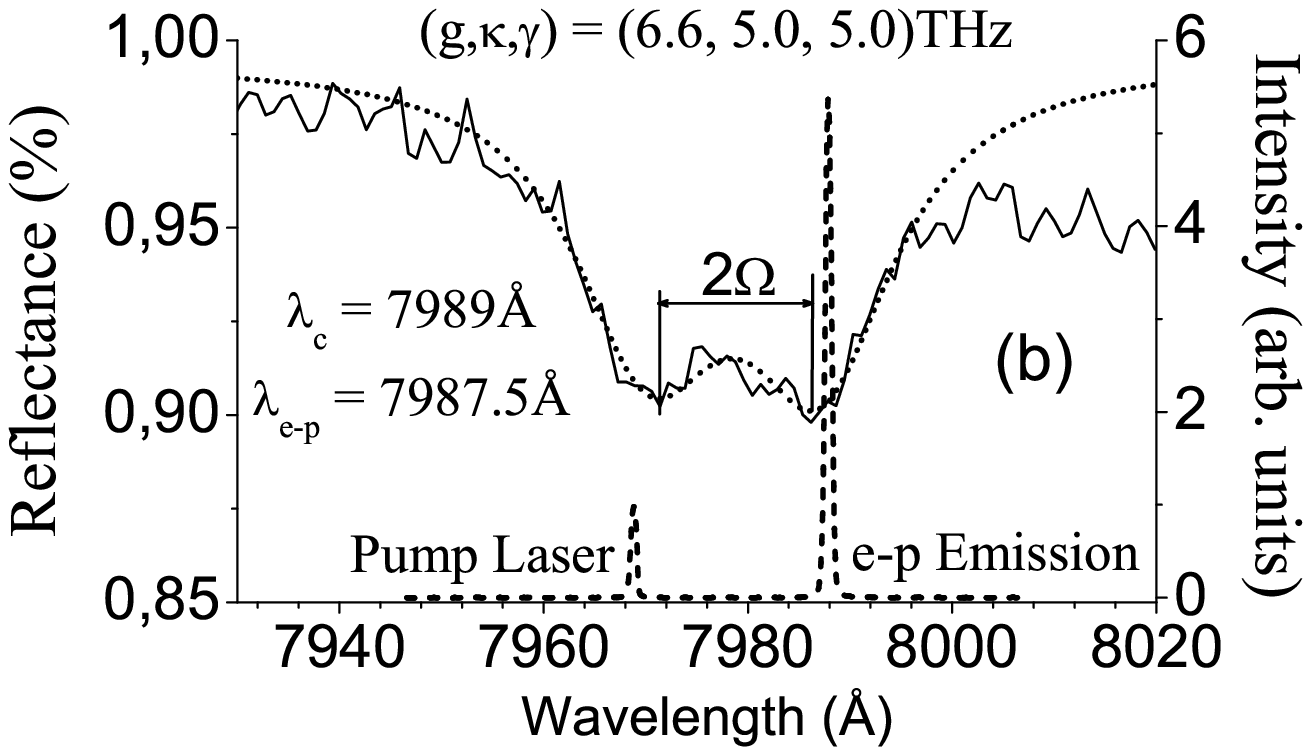}
  \includegraphics[height=5cm]{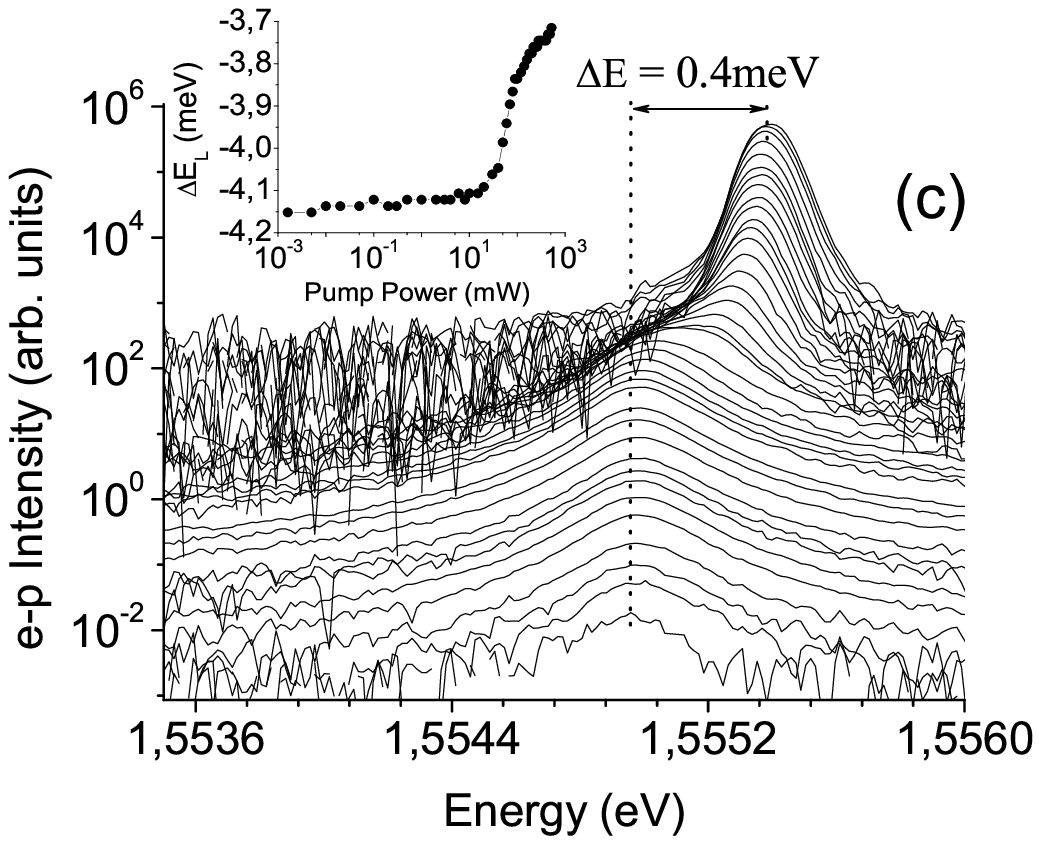}
  \caption{(a)Experimental Setup scheme. (b) The solid line (left axis) is a
  photoreflectance measurement using white light. The doted line is a fit
  using strong-coupling model\cite{Berman} ($\lambda_c$ is the
  resonance cavity wavelength, $\lambda_{e-p}$ is the exciton-polariton
  peak emission). The dashed curve (right axis) is a resonant
  photoluminescence measurement that shows the exciton-polariton
  laser emission and the laser pump. (c) Resonant photoluminescence measurement
  for several pump powers (from 1mW at 500mW). Inset: The behavior for
  the blue-shift of the e-p emission (relative peak position
  $\Delta E = E_{e-p} - E_L$, where $E_L = 1.5513eV$) vs pump
  power.}
  \label{ExpSetup}
\end{figure}

The sample\cite{Cotta} was held in a cold-finger cryostat at a
temperature of 10K. A tunable cw Ti:Sapphire laser line amplitude
was modulated using an acousto-optic modulator (AOM) coupled to a
digital function generator, producing a sinusoidal waveform pump at
1MHz (see fig.\ref{ExpSetup}-a). The lower-polartion branch (LPB) is
excited resonantly using a right circularly polarized beam
(heavy-hole excitons) through a $\lambda/4$ plate in the pump laser
path. The angle of incidence was controlled by shifting the pump
laser laterally toward a lens, being fixed in $12^o$ and focused to
a $40\mu m$ spot size. This same lens is used to collimate the light
emitted by the sample within a broad cone concentrated within $\pm
15^o$ due to the effect of the ``polariton energy trap'', formed by
the dispersion relation. The emission spectra are obtained using a
spectrometer with a grating (1800 grooves/mm) and a cooled CCD.
Detection of the pump and the e-p emission intensities were
performed using two variable-gain photodetectors. For OB
measurements, we calibrate the system splitting the input beam,
where the collected signals are coupled to X-Y oscilloscope
channels, and set the photodetector gains to achieve a straight line
relation between them.

In our discussion we will consider only the signal emission, because
it has a recombination probability much larger than idler emission,
for the used incident angle. Experimentally, a scanned detection
angle solved experiment around the $\overrightarrow{k} =
2\overrightarrow{k_p}$  direction, with the pump energy maintained
always resonant with the lower polariton branch, was realized. But,
we did not observe the idler emission in these experimental
conditions, same in the range of the detuning to the exciton-cavity
and laser-exciton used in this paper.

Reflectance measurements (see fig.\ref{ExpSetup}-b) show a Rabi
interaction frequency between excitons and photons of $\hbar\Omega
\equiv [g^2 - (\gamma - \kappa)^2/4]^{1/2} \approx 3.4meV$, where
$g$ is the rate at which the excitons are coupled with the cavity
field, $\kappa$ is the decay rate of the cavity field and $\gamma$
is the spontaneous emission rate. The resonance condition was
obtained with a detuning between the emission of the e-p mode and
the cavity mode as $\Delta_c = E_{cav} - E_{e-p} = -0.2meV$. The
detuning between the pump laser and the e-p lower polariton laser
was $\Delta_L = E_{e-p} - E_{L} = -3.7meV$.

Fig. \ref{ExpSetup}-c shows a sequence of emission spectra pointing
the repulsive polariton-polariton interaction, manifesting itself a
phase-space filling effect (as first noted in Ref.\cite{Schmitt}),
leading to bleaching of the excitonic oscillator strength and a
blue-shift of the exciton resonance. The maximum blue shift of 0.4
meV is still much less than half of the UP-LP splitting of 3.4meV;
i.e., the energy of the emission always remains distinctly below the
QW exciton and cavity photon energies. This ensures that the
polariton is still the normal mode of the system.

\begin{figure}[htbp!]
  \includegraphics[height=5cm]{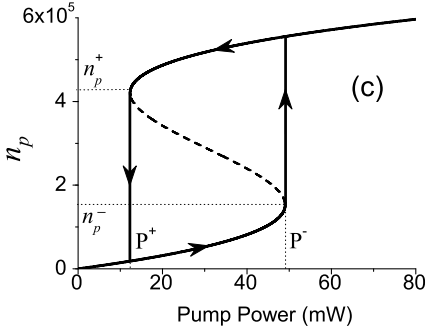}
\caption{The mean number of polaritons
  $n_p$ vs the pump power $P_{in}$. The dashed line shows the unstable
  region. The arrows indicate the hysteresis cycle obtained by scanning
  the input power in both directions.}
  \label{TheoreticalOB}
\end{figure}

Using the theoretical results discussed in Ref.\cite{Giacobino} and
$a_{exc} = 100\AA$ (for 2D Bohr radius of the exciton in
GaAs\cite{Wolfe}) we obtain the theoretical OB curve shown in
fig.\ref{TheoreticalOB}, and the bistability condition: $\Delta_L <
-\sqrt{3}\Gamma_{e-p} = -1.3meV$, where $\Gamma_{e-p}$ is the
exciton-polariton linewidth obtained from cavity ($\Gamma_c$),
exciton ($\Gamma_{exc}$) linewidths and the Hopfield
coefficients\cite{Hopfield}. In our sample, the cavity quality
factor $Q = E_c/\Gamma_c = 1.5533eV/1.04meV \simeq 1500$ and
$\Gamma_{exc}= 0.11meV$, resulting $\Gamma_{e-p} = 0.75meV$.

A theoretical analysis shows that $n_p^+ \approx 2.8n_p^-$ and $P^-
\approx 4P^+$, where $n_p^\pm$ and $P^\pm$ (defined in the
fig.\ref{TheoreticalOB}) are respectively the threshold for mean
polariton number and pump power to the bistability forward ($-$) and
turning ($+$) points. The minimum exciton-laser detuning and its
respective pump power threshold is obtained when $\Delta_L^{min} =
-\sqrt{3}\Gamma_{e-p} = -1.3meV$. With this condition, we find the
minimum cavity detuning $\Delta_c^{min} = [\hbar\Omega(2\Gamma_c -
\Gamma_{exc})/(2\sqrt{2\Gamma_c\Gamma_{exc}})] = -2.4meV$ (using the
mentioned parameters).

Thus, the model predicts a dispersive OB due to e-p population in
our sample, but these theoretical results, and an analysis of the
threshold equations for cavity-exciton and exciton-laser detuning
using ref.\cite{Giacobino} do not foresee a crossing effect between
the branches in the OB curve.

\begin{figure}[htbp!]
  \begin{center}
  \includegraphics[height=3.5cm]{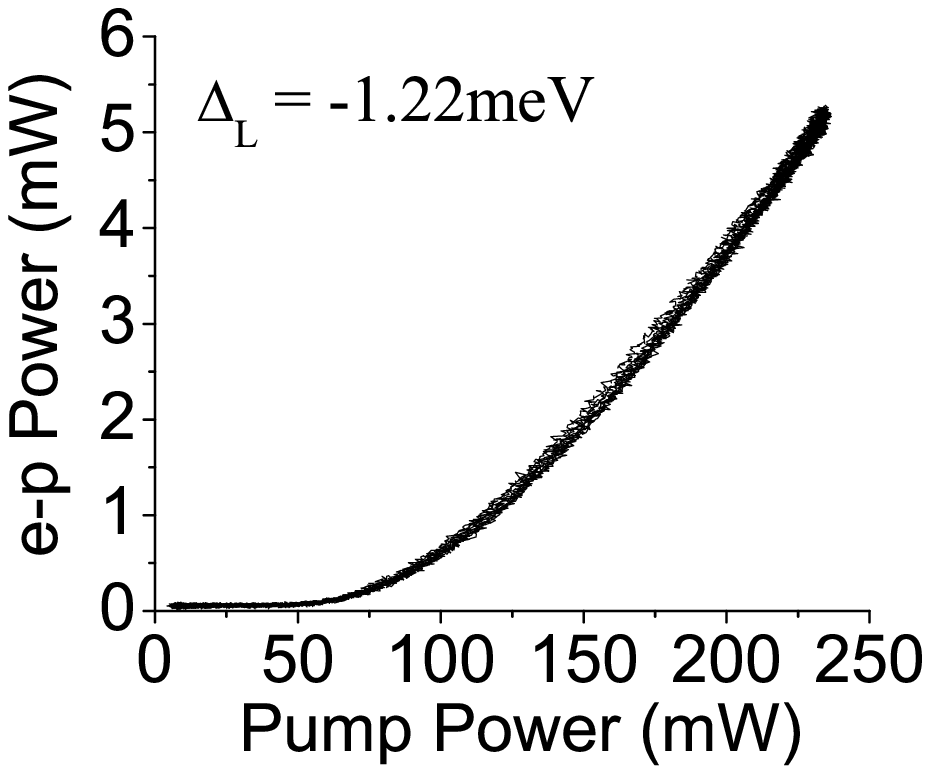}
  \includegraphics[height=3.5cm]{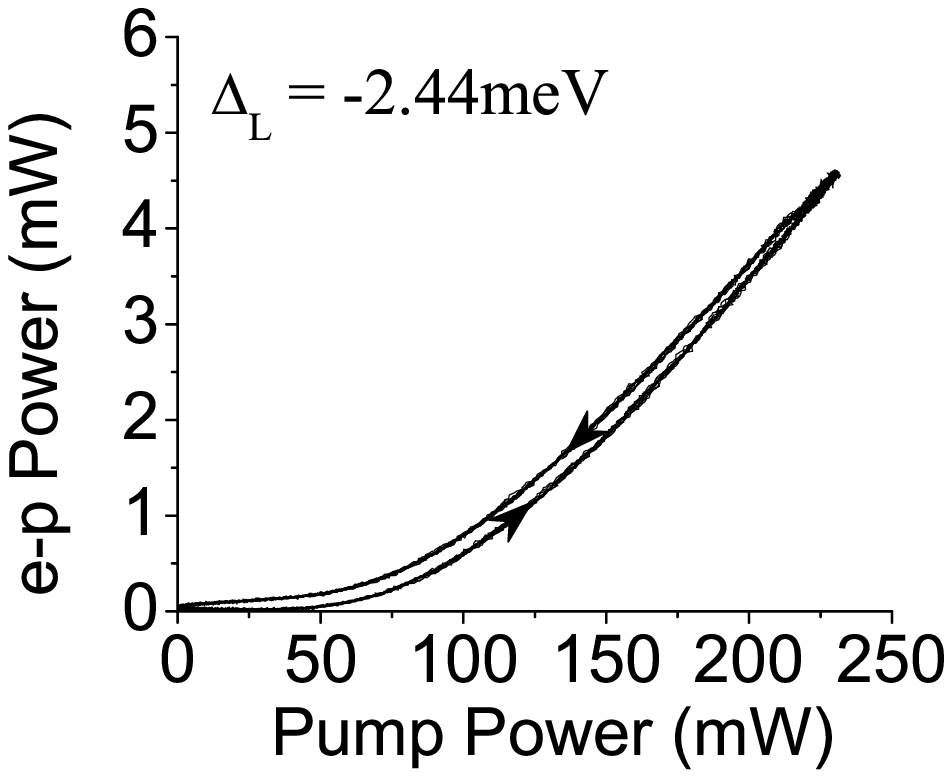}
  \includegraphics[height=3.5cm]{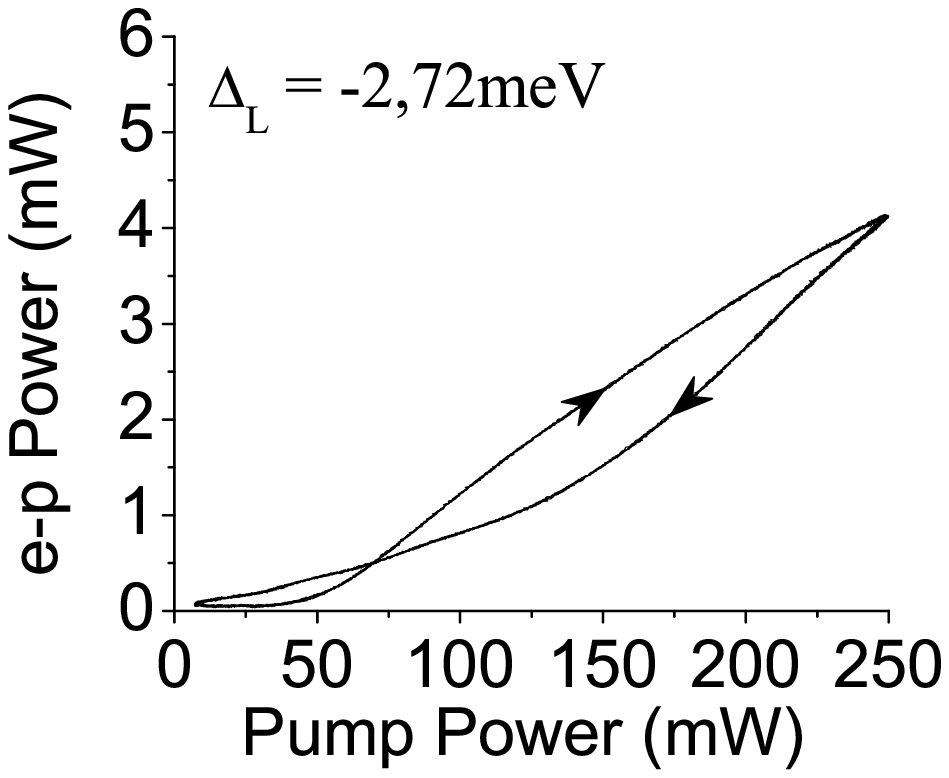}
  \includegraphics[height=3.5cm]{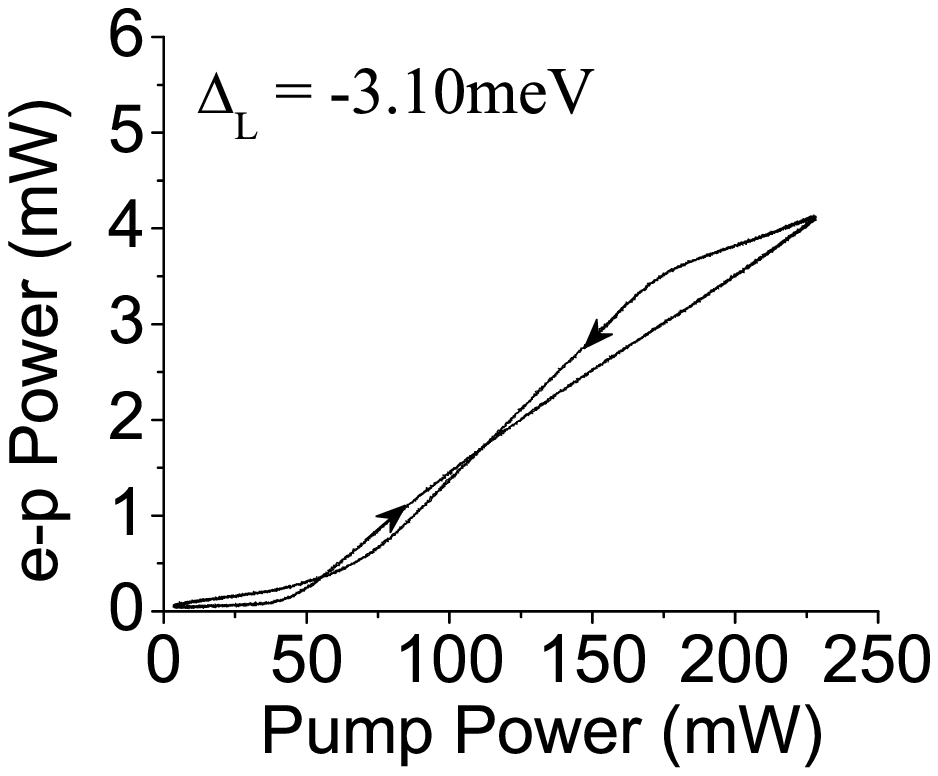}
  \end{center}
  \vspace{-0.5cm}
  \caption{Bistability sequence curves for
  a fixed cavity detuning ($\Delta_c = -2.59meV$) and several pump
  laser detuning. The e-p power emission error is $3\%$.}
  \label{ExpOB}
\end{figure}

Experimental results for the OB are shown in fig.\ref{ExpOB} for
different excitation energies, where one can see two crossings
between the turn-on and the turn-off curves for $\Delta_L =
-3.10meV$. The bistable behaviors can be of dispersive or absorptive
nature, or a competition between both. Moreover, since thermal
conduction plays an important role in the dynamics of thermal
optical bistable devices, an expression for the thermal conduction
time is useful. If a laser beam suddenly heats a cylinder of length
much smaller than the beam radius $r_o$, the heat diffusion equation
leads to a thermal conduction time of\cite{Hyatt}
\begin{equation}
\tau_{th} = \frac{c_v \rho r_o^2}{(2.4)^2 K_c}
\end{equation}
where $c_v$ is the specific heat, $\rho$ is the density, and $K_c$
is the thermal conductivity. $K_c$ and $c_v$ both depend upon
temperature, so that $\tau_{th}$ can change by several orders of
magnitude. Also, $\tau_{th}$ is proportional to the square of the
pump laser beam radius; small dimensions can result in short
conduction times. For GaAs at $10K$, $c_v =
2.7J/(kg.K)$\cite{SadaoAdachi}, $K_c = 1400 W/(m.K)$, $\rho =
5316.5kg/m^3$ (see Ref.\cite{Peaker}) leading to $\tau_{th} \approx
712 ps$. For a modulated pumping frequency of 1MHz we have an quasi
steady-state, in which the input is turned on and off through
sinusoidal pulses with duration much larger than $\tau_{th}$, but
shorter than the time to reach complete thermal equilibrium with the
surroundings. Thus, the system can show an overshoot in the
switch-on for $n_p$, soon after the input reaches it is maximum.
Although the input is reduced as quickly as it is increased, we also
see an overshoot in switching off for the cooling operation.

For thermal bistability, the effective pump power per pass that
excites the gain medium in the laser cavity is: $P_p =
P_oTR\times(1-L_i)^2\times(1-\exp[-\alpha_c L_c])^2\times
\exp[-2\alpha_{qw} L_{qw}] = 16 mW$, where $P_o = 240mW$ is the pump
power at the sample surface, $\alpha_{c(qw)} = 13267cm^{-1} (1.5
\times 10^5cm^{-1})$ is the optical absorption coefficient of $GaAs
(Al_{0.3}Ga_{0.7}As)$ (at $10K$ and excited
resonantly)\cite{SadaoAdachi}, $L_i$ is internal cavity loss, where
$L_i = \gamma \tau_i$ ($\gamma$ is the internal cavity decay
obtained in fig.\ref{ExpSetup}-b and $\tau_i$ is the cavity round
trip time), $L_{c(qw)}$ is cavity (quantum well) length, $T = 8.5\%$
and $R = 90\%$ are the cavity transmission and reflectance at
resonance, respectively. For $n_i= \tau_c/\tau_i \approx 400$
internal reflections (the cavity lifetime is $\tau_c = Q/2\pi\nu_c
\approx 0.38 ps$), $P_p \simeq 70mW$. But only a part of this
excitation power is used to heat the sample ($P_{therm}$); the rest
is used to produce cavity photons ($P_{ph}$), so that $P_p =
P_{therm} + P_{ph}$. To obtain $P_{therm}$ we analyze the e-p
dispersion curve, finding $P_{therm} = P_p (1 - E_{e-p}/E_L) = 170
\mu W$, where $E_{e-p}$ and $E_L$ are exciton-polariton peak
emission and pump laser energies, respectively. Using the fact that
each exciton is scattered by one-phonon to bottom of the LPB, we
have $N = P_{ph}\lambda_L\tau_{sp}/(hc)$ as the mean phonon number
($\tau_{sp} = 3ns$ is the excitonic radiative spontaneous
lifetime\cite{Yamamoto}, $hc/\lambda_L$ is the energy of each pumped
photon laser). Thus, during time $\tau_{th}$, the temperature at the
cavity can be raised by $T = P_{therm}\tau_{th}/(NK) \simeq 10K$.

\begin{figure}[htbp!]
\begin{center}
  \includegraphics[height=3.5cm]{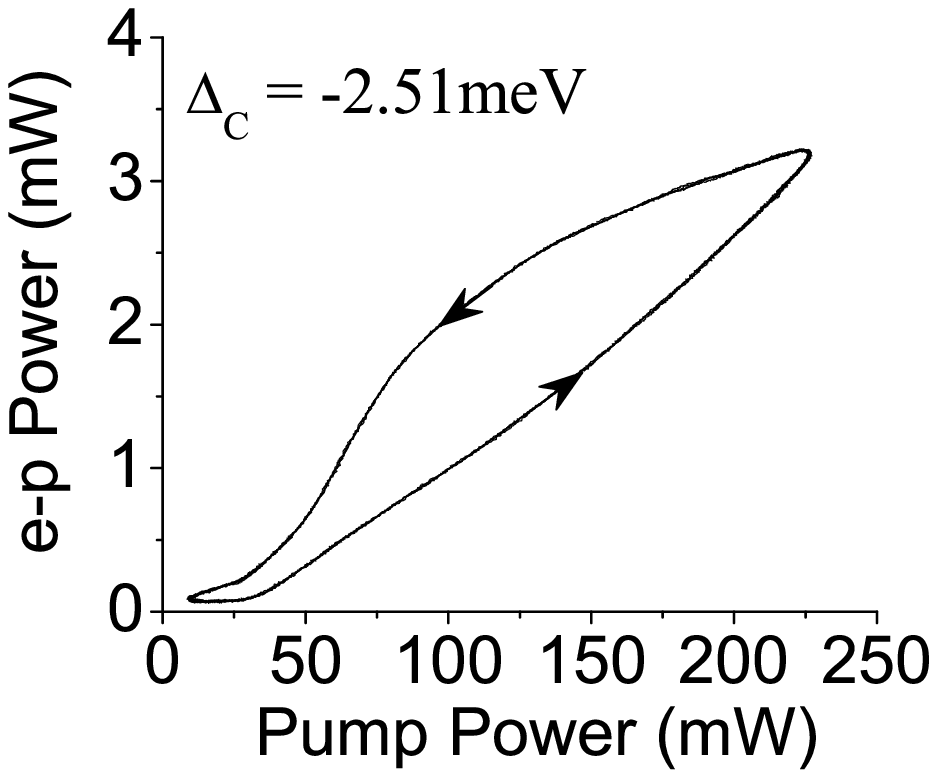}
  \includegraphics[height=3.5cm]{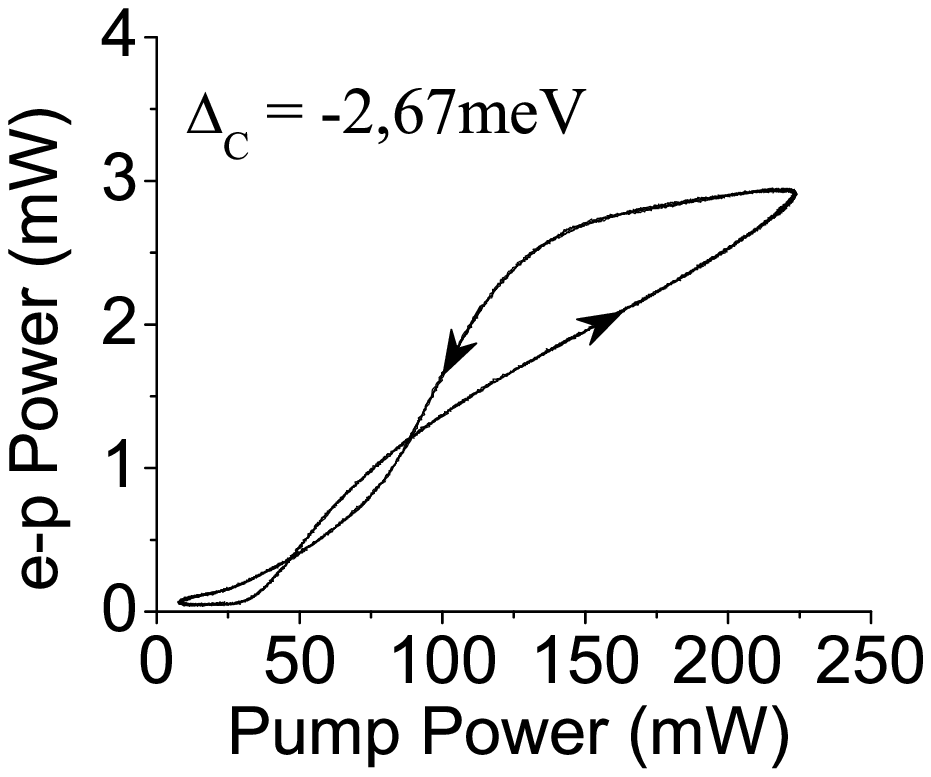}
  \includegraphics[height=3.5cm]{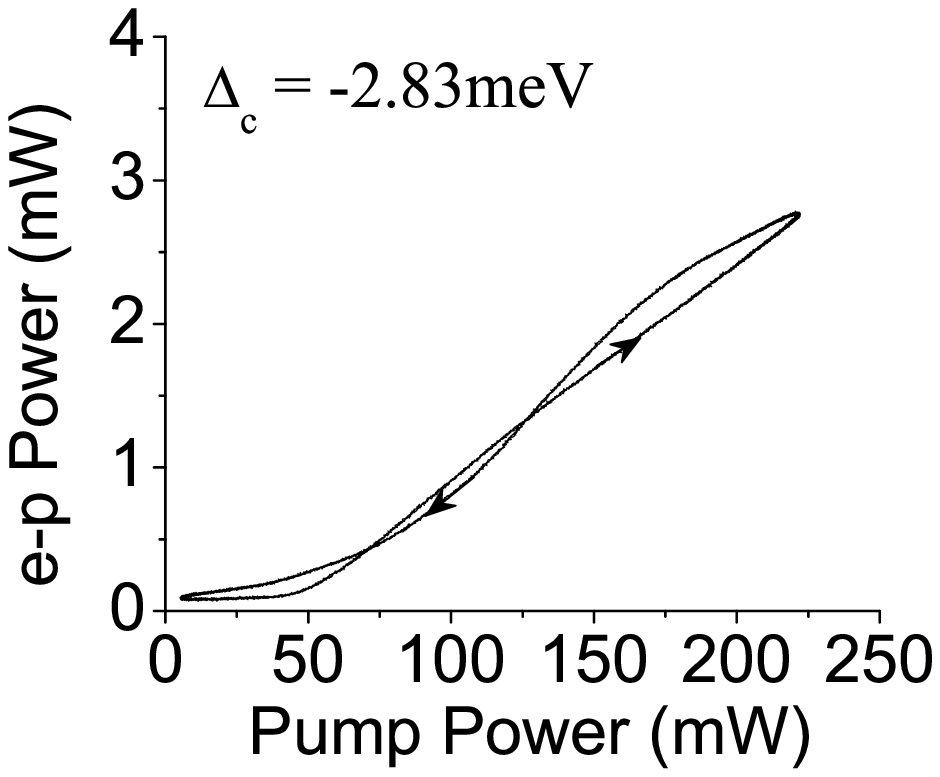}
  \includegraphics[height=3.5cm]{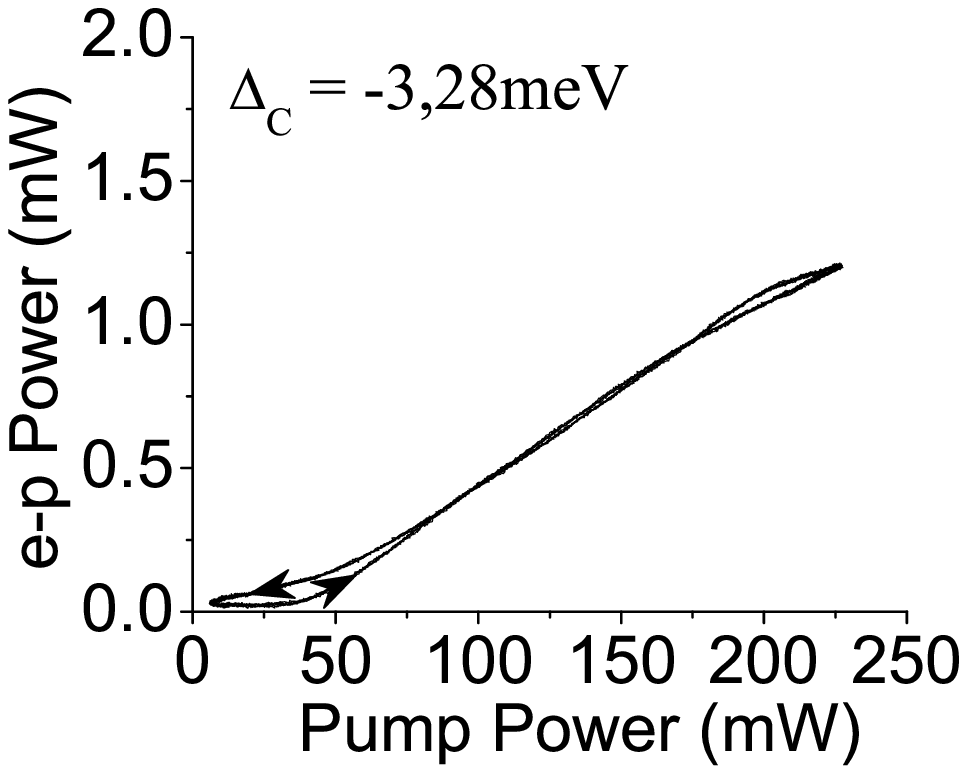}
\end{center}
\vspace{-0.5cm}
  \caption{Bistability sequence curves for a fixed laser energy
  ($\Delta_L = -3.09meV$) and several cavity detuning. The e-p power emission error is $3\%$.}
  \label{CavDetuning}
\end{figure}

The excitonic effect provides a change of $\Delta n_2/I = -0.005
cm^2/kW$ and the thermal effect $dn/dT = +0.0004 K^{-1}$ (see
Ref.\cite{Hyatt}). Thus, in the first case, the changes in
refractive index produce a deformation of $\Delta L_{QW} = -2.8
\AA$, while in the second case the heating results in an expansion
of the cavity, so that we have both a cavity refractive index change
$\Delta n$ (temperature-dispersive effect) and a cavity length
change $\Delta L$ (temperature-expansive effect), resulting in
$\Delta L_c = L\Delta n + n\Delta L = 10\AA$. These results show
that it is possible to observe a ``backward'' hysteresis due to
thermal effects, because of the increase in cavity optical path with
temperature, where the switch-down threshold intensity is higher
than that for switch-up. When the pump power reaches its maximum,
the entire cavity is heated, but the microcavity is already in the
e-p laser regime, and the dispersive OB due to e-p population is
predominant. This heating (some fraction of a microsecond later),
affords a thermal-dispersive effect that achieves values, for the OB
effect, greater than previously cited and the first crossing occurs.
But, this effect is decreased because of the dissipation of the heat
inasmuch as the pump power is reduced. Both heating and cooling
processes occur above the e-p laser threshold, because the
dissipation time is faster than the modulation of the pumping laser.
Thus, as soon as the dissipation of heat is completed and a second
crossing occurs.

Clearly, this thermal optical path-length mechanism is not a second
order process in the rigourous sense. That is, the index does not
depend upon the instantaneous value of the intensity according to $n
= n_o + n_2I$, but rather on the time that the heat is dissipated by
the sample. Moreover, the optical path length change depends on the
total temperature variation, i.e., it depends on competition between
absorption and heat conduction. Roughly speaking, it depends upon
the energy absorbed during the last thermal conduction time. Thus,
the variations in the refractive index is position-dependent, and
should be weighted by the local optical intensity due to cavity
resonance. These variations in the losses and/or in the absorption
of the gain medium change the local heating of the cavity and the
time to heat the lattice ($\tau_{th}$) as even. In the
fig.\ref{CavDetuning} we can see a variation of this heat
dissipation time for a fixed laser-exciton detuning, observing the
curve-contour.

Summarizing, we find both single and double crossings in the OB
curve due to a competition between the dispersive Kerr-like effect
(caused by e-p population) and thermal effect, governed by the heat
dissipation time. An analysis shows that the thermal dependence has
a strong weight in the competition of the bistability effect, mainly
due to temperature-dispersive effect. In spite of the temperature of
the sample achieve $\sim 20K$, the system remains in the strong
coupling regime, since reflectance measurements showed it at
temperatures until $40K$\cite{Cotta}.

The authors are very grateful to CNPq and FAPEMIG by financial
support and Carlos Henrique Monken, Ronald Dickman and Alfredo
Gontijo for stimulating discussions and helpful advises.

\end{document}